\documentclass{amsart}
\usepackage{amsfonts,amsmath,amssymb}
\usepackage[autostyle=true]{csquotes}
\usepackage{tikz,pgfplots}
\usepackage{url}
\usepackage{soul,xcolor}
\usepackage{graphicx}
\usepackage{lineno}
\usepackage{placeins}
\usepackage{enumitem}
\usepackage{nicefrac}
\usepackage{booktabs}
\PassOptionsToPackage{dvipsnames,svgnames}{xcolor}     
\usepackage{xkeyval,tkz-base}
\usetikzlibrary{arrows,calc,3d,hobby}
 \makeatletter%

\usepackage{bm}
\usepackage{caption}
\usepackage{subcaption}

\begin{document}

\setstcolor{red}



\newcommand{\Einc}{E^{\rm inc}}
\newcommand{\Esca}{E^{\rm sca}}
\newcommand{\Etot}{E^{\rm tot}}
\newcommand{\EL}{E_L}
\newcommand{\x}{\bm{x}}
\newcommand{\y}{\bm{y}}
\newcommand{\Rtwo}{\bm{R}^2}

\title[PNJs as power flux funnels]{Photonic nanojets as emergent free-space\\power flux funnels}
\author[Karamehmedovi\'c]{Mirza Karamehmedovi\'c$^{1,\ast}$}
\author[Placinta]{Cristian Placinta$^{1}$}
\author[Mikkelsen]{Tobias Abilock Mikkelsen$^{1}$}
\author[Gl\"uckstad]{Jesper Gl\"uckstad$^{2}$}
\thanks{\flushleft$^{\ast}$mika@dtu.dk\\$^{1}$Department of Applied Mathematics and Computer Science, Technical University of Denmark, DK-2800 Kgs. Lyngby, Denmark.\\$^{2}$SDU Centre for Photonics Engineering, University of Southern Denmark, DK-5230 Odense, Denmark.}

\begin{abstract}
A reduced local field model derived from full-wave electromagnetic simulations shows that photonic nanojet formation corresponds to an emergent mesoscopic funnel of propagating power flux sustained by an effective free-space transverse mode structure. This interpretation moves beyond purely geometric-optics or interference-based explanations by identifying a self-consistent redistribution of phase gradients and effective longitudinal wavenumber near the nanojet waist. The model quantitatively captures characteristic nanojet morphology, including the formation and local structure of the jet waist. It also yields a geometry-independent lower bound on the nanojet waist, linking transverse confinement to the effective axial wavenumber through an explicit trade-off. The model establishes a direct connection between full-wave Maxwell fields and a reduced free-space oscillator description, yielding new physical insight into nanojet confinement and suggesting design principles for nanojet-assisted imaging, lithography, and subwavelength field localization.
\end{abstract}

\maketitle

\section{Introduction}

Photonic nanojets (PNJs) - highly localized, high-intensity beams forming in the near field of transparent dielectric micro-elements - were first observed experimentally in 2006~\cite{Heifetz}. Their ability to generate sub-diffraction features without relying on cavity resonances or complex material platforms has stimulated broad interest in applications ranging from label-free microscopy and nanoparticle manipulation to super-resolution metrology~\cite{Darafsheh-2021,Huszka}; see also~\cite{2023-phase-only_PNJ} for a more comprehensive review of the literature.

Despite extensive numerical and experimental investigation, the physical mechanism governing PNJ formation and confinement remains incompletely understood. A central difficulty is that PNJs are inherently mesoscopic optical structures, extending over only a few to a few tens of wavelengths. In this regime, geometric-optics constructions are invalid, while full-wave simulations, though capable of reproducing PNJs with high fidelity, offer limited physical insight: the jet appears in the computed field, but its origin, internal structure, and sensitivity to material and geometric parameters remain unclear. As a result, much of the existing theoretical understanding of PNJs relies on special-case expansions or numerical decompositions that lack generality.

Several influential explanations have nevertheless been proposed. Itagi and Challener~\cite{Itagi-2005} reformulated the Mie series for a two-dimensional circular micro-element into a Debye-type modal expansion, identifying a doubly refracted cylindrical component as the dominant contributor to the jet. However, this interpretation is based on numerical inspection and does not explain why this mode should dominate beyond highly symmetric geometries. Devilez et al.~\cite{Devilez-2008} and Geints et al.~\cite{Geints-2012} analyzed Lorenz–Mie expansions for spherical elements and attributed PNJ formation to low-order propagating components and internal resonances, again supported primarily by numerical evidence in symmetric settings. Geometric-optics–based approaches~\cite{Lecler-2019} provide qualitative intuition, but rest on asymptotic assumptions that do not apply at mesoscopic scales. Overall, existing PNJ models remain largely computational and case-specific, with no general principle explaining confinement or enabling systematic control.

In this work, we introduce a mesoscopic, mode-theoretic model of photonic nanojet formation based on a local amplitude–phase reduction of the Helmholtz equation in the nanojet waist region. The resulting reduced model identifies a universal phase geometry that produces an effective transverse confining potential in free space. By analyzing numerically computed phase fields in both two- and three-dimensional configurations, we observe that the transverse phase curvature varies linearly with the axial coordinate and vanishes at a finite axial position. This produces a characteristic “transverse phase hourglass”, or "PNJ phase funnel," described by a simple polynomial expression. This phase geometry reflects a general mechanism by which a local reduction of the axial phase gradient generates an effective transverse confining potential in free space. The resulting local free-space oscillator picture provides a new interpretation of PNJ formation that goes beyond conventional descriptions based on diffraction, interference, or mesoscale focusing, and leads to explicit constraints linking transverse confinement to the effective axial phase gradient.

We describe the PNJ phase funnel in Section~\ref{sec:trap} and use it to derive the free-space oscillator model in Section~\ref{sec:qqo}. Numerical validation is presented in Section~\ref{sec:numerical}, followed by conclusions in Section~\ref{sec:conclusion}.

\section{The PNJ phase as a power flux funnel}\label{sec:trap}

In this section, we present observations of the phase associated with numerically computed photonic nanojet (PNJ) fields. With the time-harmonic convention $e^{j\omega t}$ used throughout this paper, our central observation is that, in the vicinity of a PNJ, the phase function $\varphi$ admits the approximation
\begin{equation}\label{eqn:ptm}
\varphi(\bm x)\approx\varphi_{\rm funnel}(\bm x)=\varphi_0-k_{\rm eff}\bm x\cdot\widehat{\bm d}-\frac{\Omega}{2}(\bm x\cdot\widehat{\bm t})^2(\bm x-\bm x_{\rm PNJ})\cdot\widehat{\bm d},
\end{equation}
for a time-harmonic PNJ field with temporal dependence $e^{j\omega t}$ that propagates along the unit vector $\widehat{\bm d}$, features the PNJ neck at $\bm x_{\rm PNJ}$, and is approximately axially symmetric near $\bm x_{\rm PNJ}$. Here, $\widehat{\bm t}$ is a unit vector in the transverse direction, that is, orthogonal to $\widehat{\bm d}$, $\varphi_0$ is a constant phase offset, and $k_{\rm eff}<k_0$ is an effective axial wavenumber, with $k_0=2\pi/\lambda_0$ the free-space wavenumber at wavelength $\lambda_0$. The parameter $\Omega$ characterizes the transverse confinement strength; its significance will become clear in Section~\ref{sec:qqo}. The reason we interpret~\eqref{eqn:ptm} as a 'power flux funnel' is that, for the fields considered in this work, the time-average propagating power flux density vector $\nicefrac{1}{2}\Re(\bm E\times\bm H^{\ast})$ is directed along the negative of the phase gradient; we shall namely factorize the (dominant) electric field component into amplitude and phase, $E_{\rm dom}=ae^{j\varphi}$, and using this in conjunction with the Maxwell-Faraday equation $\nabla\times\bm E=-j\omega\mu\bm H$ gives the complex Poynting vector, under a locally scalar dominant-field approximation, as
\begin{equation*}
\bm S=\tfrac12\bm E\times\bm H^{\ast}\approx\frac{1}{2j\omega\mu}E_{\rm dom}\nabla E_{\rm dom}^{\ast}=-\frac{a}{2\omega\mu}\left(a\nabla\varphi+j\nabla a\right).
\end{equation*}
Specifically, the propagating (active) power flow $\Re\bm S$ is along the negative phase gradient, while the reactive power flow $\Im\bm S$ is along the negative amplitude gradient. Now
\[
-\nabla\varphi_{\rm funnel}(\bm x)=\left(k_{\rm eff}+\frac{\Omega}{2}(\bm x\cdot\widehat{\bm t})^2\right)\widehat{\bm d}+\Omega((\bm x-\bm x_{\rm PNJ})\cdot\widehat{\bm d})(\bm x\cdot\widehat{\bm t})\widehat{\bm t},
\]
so, in addition to the steady $\widehat{\bm d}$-directed (axial) power flux component, the transverse power flux is towards the propagation axis upstream of the PNJ, with $(\bm x-\bm x_{\rm PNJ})\cdot\widehat{\bm d}<0$, and away from the propagation axis downstream of the PNJ, with $(\bm x-\bm x_{\rm PNJ})\cdot\widehat{\bm d}>0$. In the simple case $\widehat{\bm d}=-\widehat{\bm z}$, we can rewrite~\eqref{eqn:ptm} in terms of
\begin{equation}\label{eqn:phi}
\varphi(r,z)\approx\varphi_{\rm funnel}(r,z)=\varphi_0+k_{\rm eff}z+\frac{\Omega}{2}r^2(z-z_{\rm PNJ}),
\end{equation}
where $r$ is the coordinate transverse to the direction of PNJ propagation, and $z_{\rm PNJ}$ is the axial location of the PNJ waist. This phase structure forms the basis for the free-space oscillator model of PNJ formation developed in Section~\ref{sec:qqo}.

\paragraph{2D case}

\begin{figure}
\centering
\includegraphics[width=0.5\linewidth]{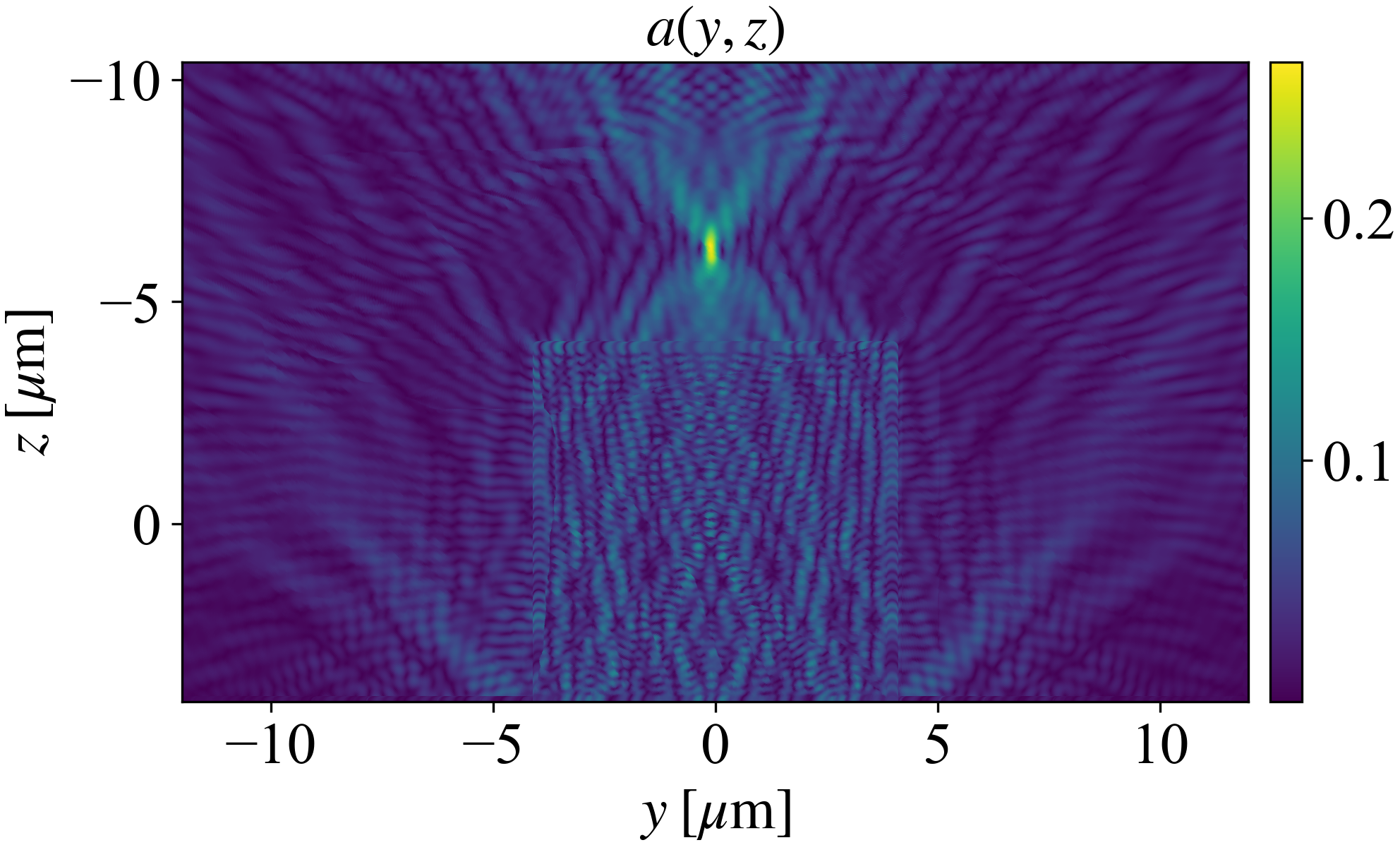}\includegraphics[width=0.5\linewidth]{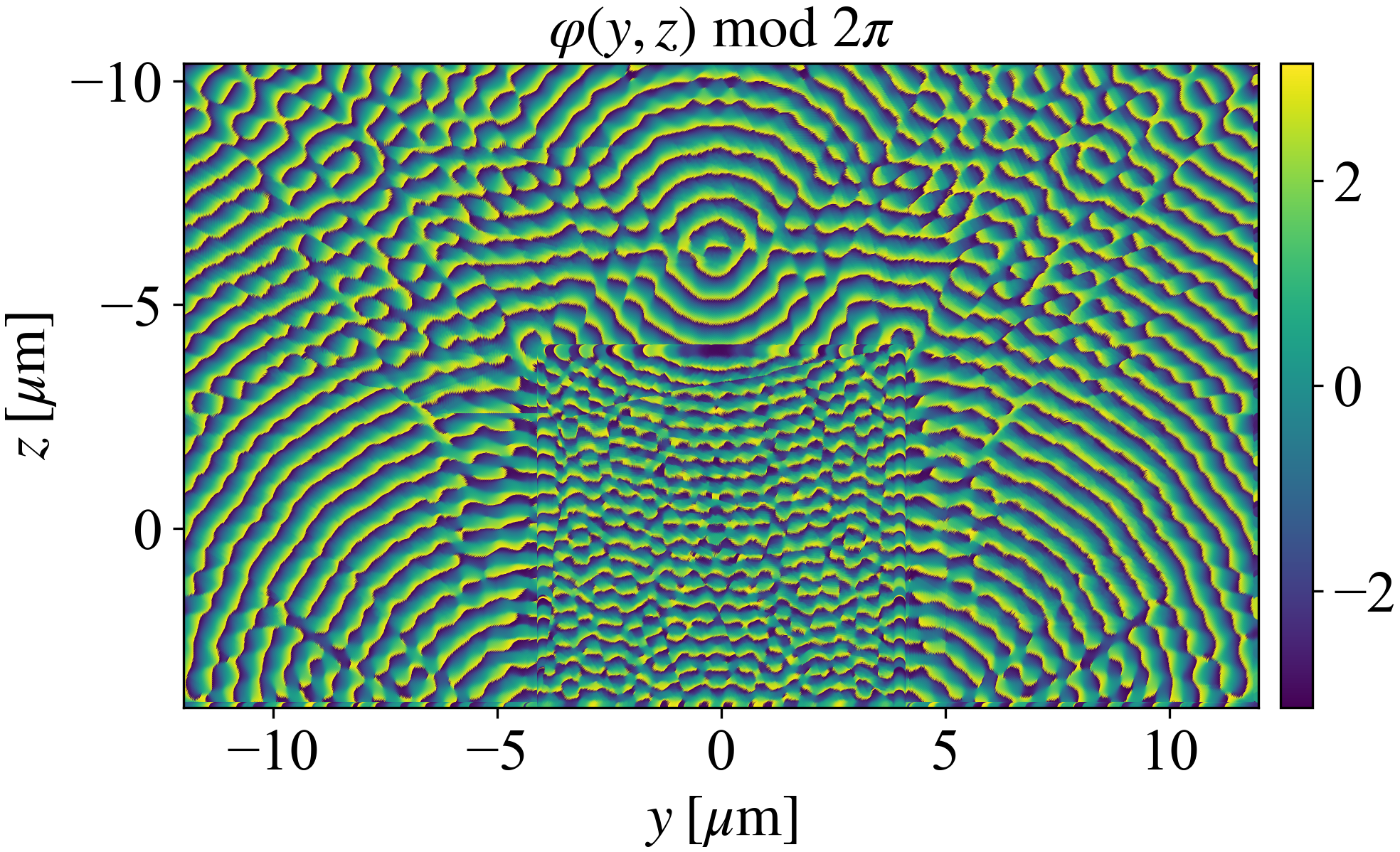}\\\includegraphics[width=0.6\linewidth]{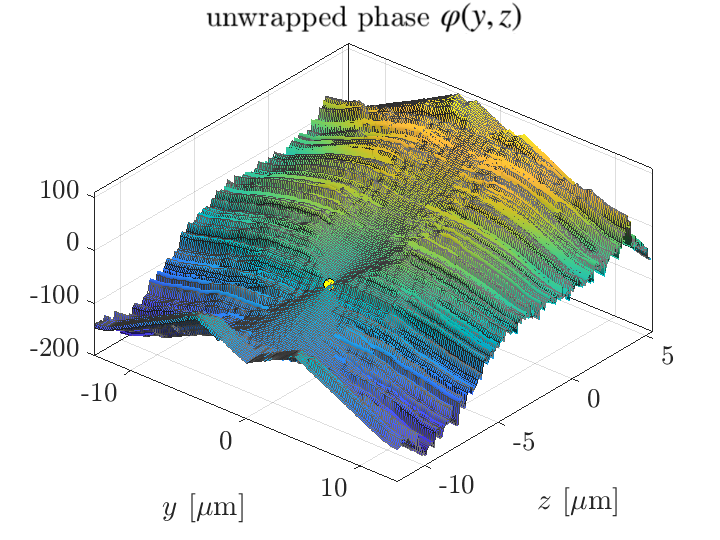}\includegraphics[scale=0.4]{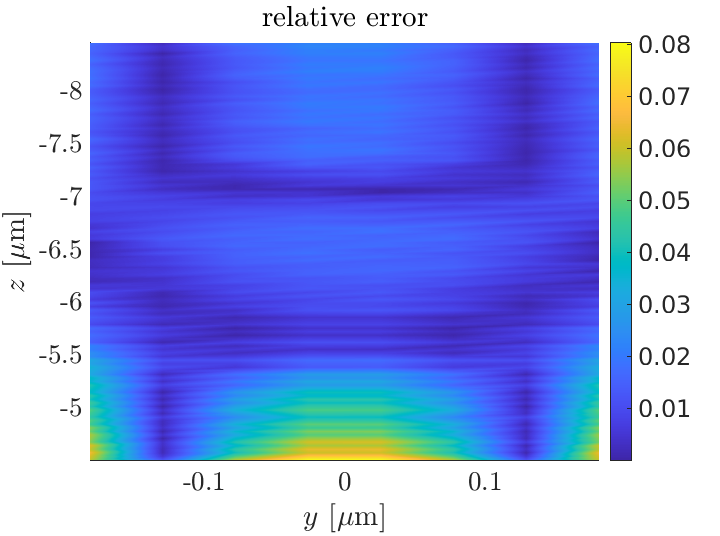}
\caption{Top: Amplitude (V/m) and phase (rad) of a TM$^x$-polarized, $-z$-directed 2D PNJ field produced by a square cross-section micro-element illuminated by structured light~\cite{2022-PNJ1}. Bottom: Unwrapped phase (rad) of the PNJ field showing the hourglass structure, and with the PNJ position indicated by a yellow dot; next, the relative error of our PNJ phase funnel model~\eqref{eqn:phi}. Mean relative error over the domain shown at bottom right is 1.5\%.}\label{fig:2DrectTM}
\end{figure}

\begin{figure}
\centering
\includegraphics[width=0.52\linewidth]{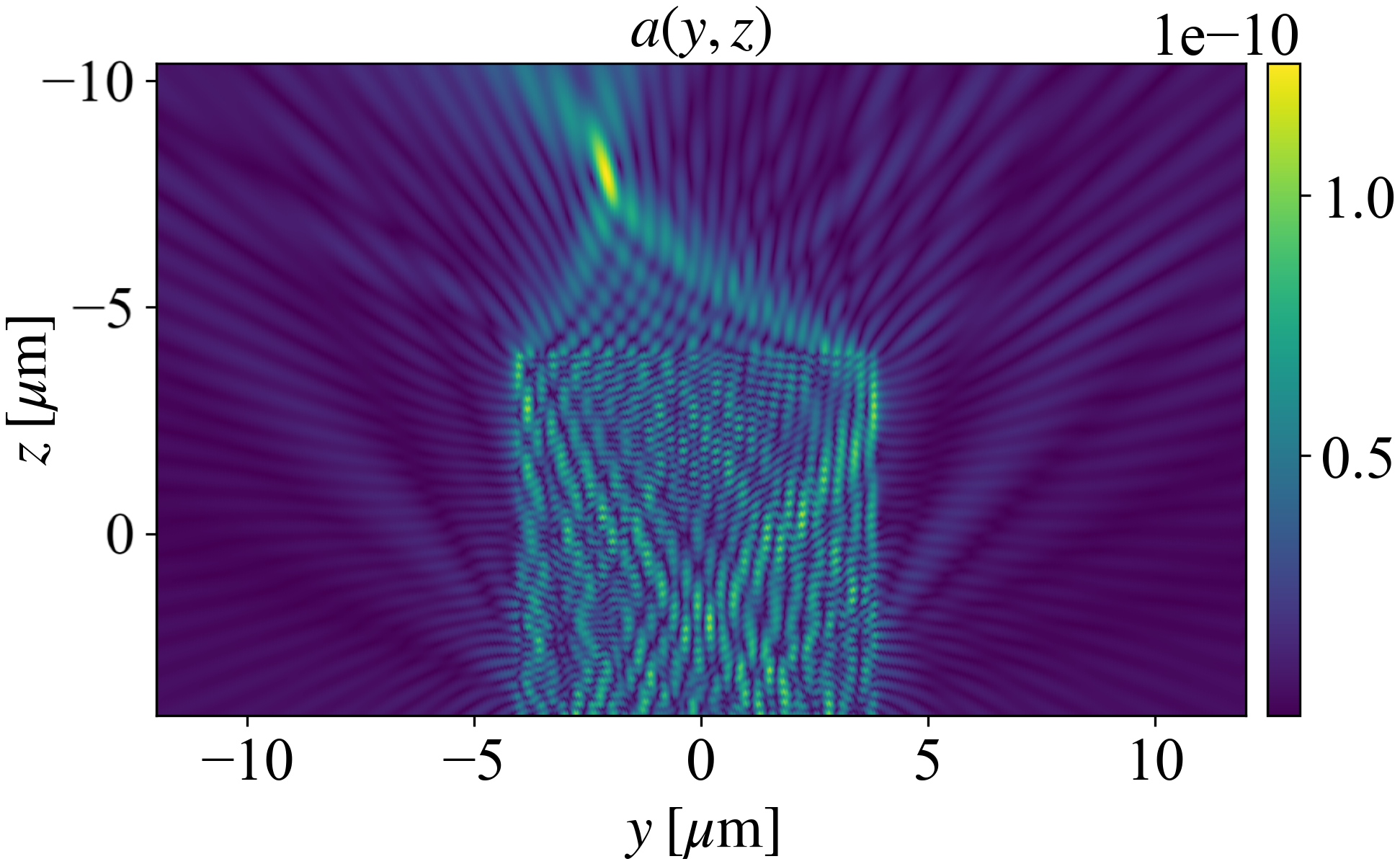}\includegraphics[width=0.5\linewidth]{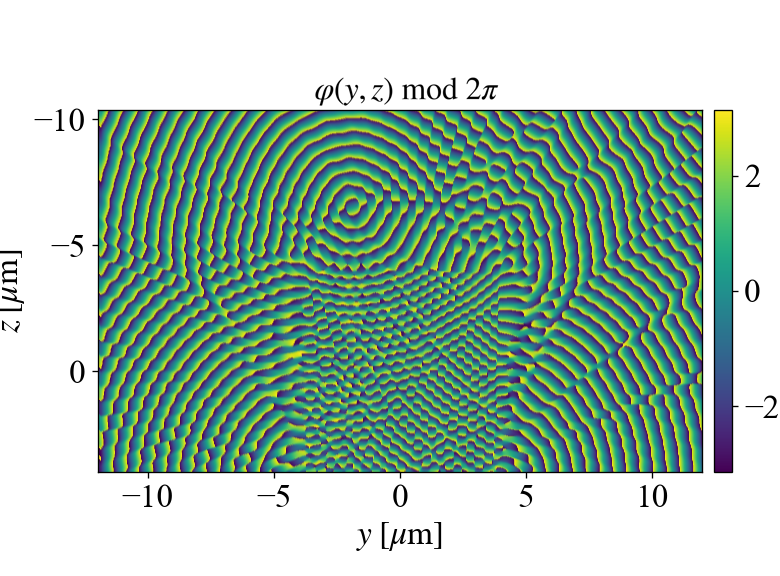}\\\includegraphics[width=0.4\linewidth]{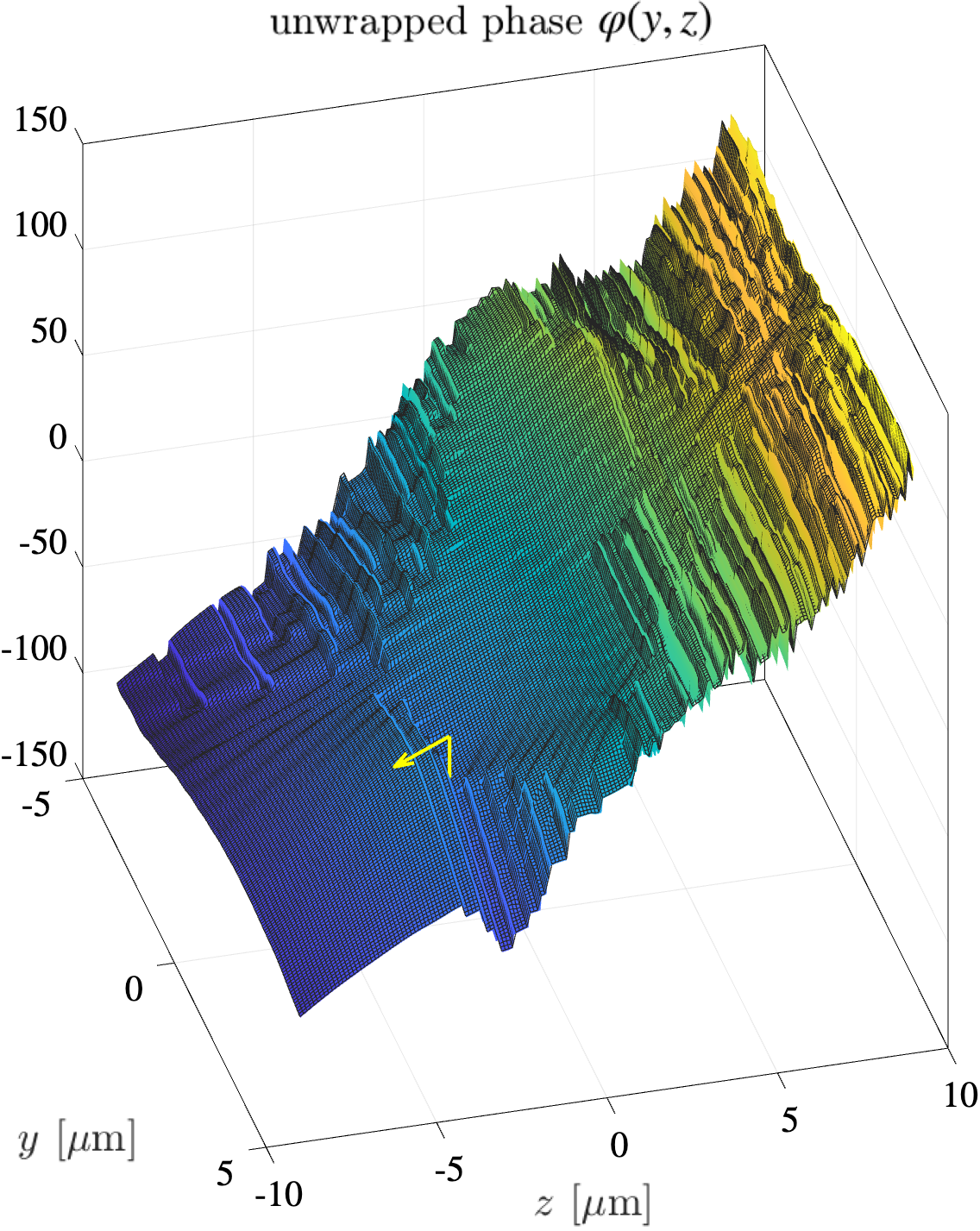}\includegraphics[width=0.5\linewidth]{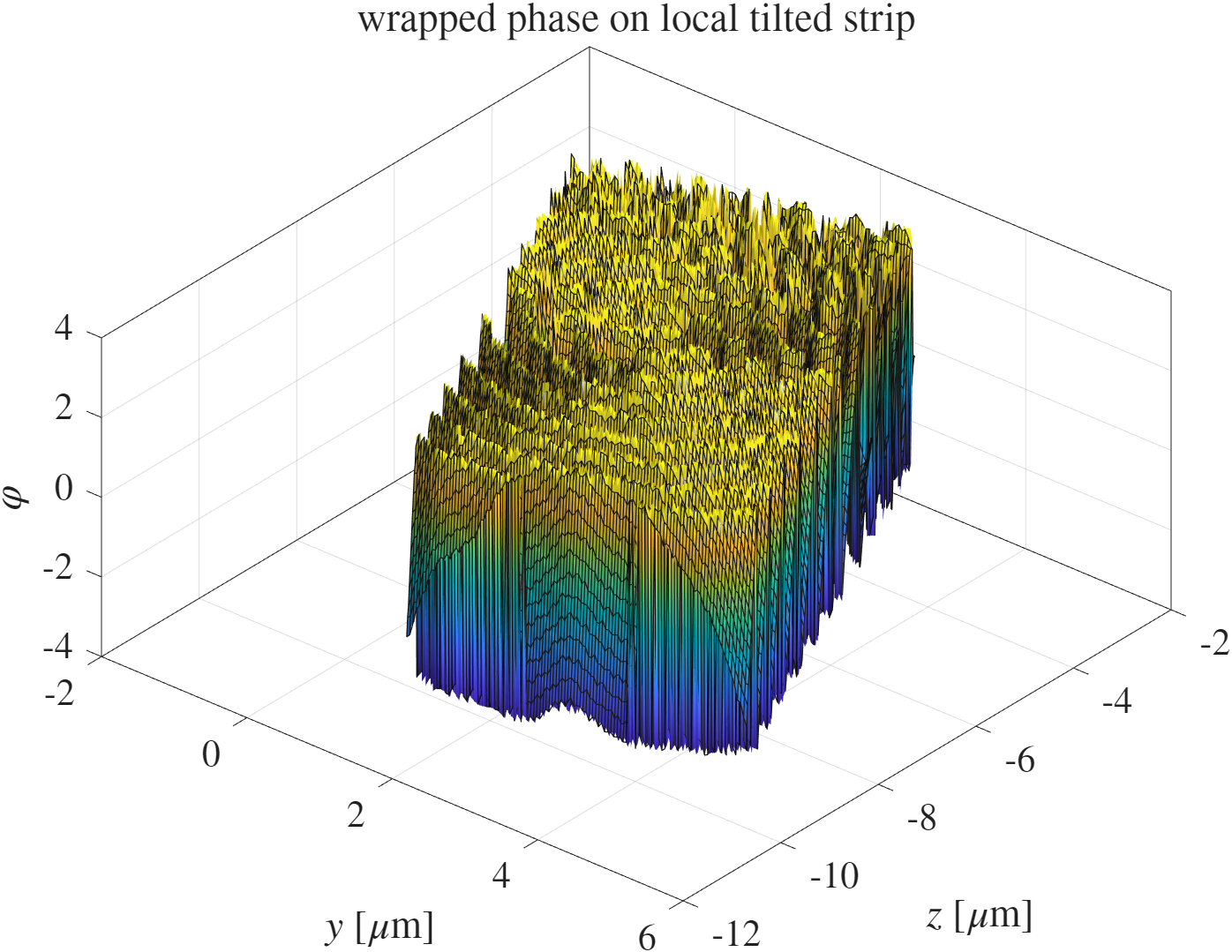}
\caption{Top: Amplitude and phase of a TE$^x$-polarized, $-z$-directed 2D PNJ field produced by a square cross-section micro-element illuminated by structured light~\cite{PNJ_CriTob}. Bottom: Unwrapped phase (rad) of the PNJ field (with visible hourglass shape, and with the PNJ position and propagation direction indicated by a yellow arrow) and the wrapped phase (rad) over a small tilted strip near the PNJ neck.}\label{fig:2DrectTE}
\end{figure}

\begin{figure}
\includegraphics[width=0.35\linewidth]{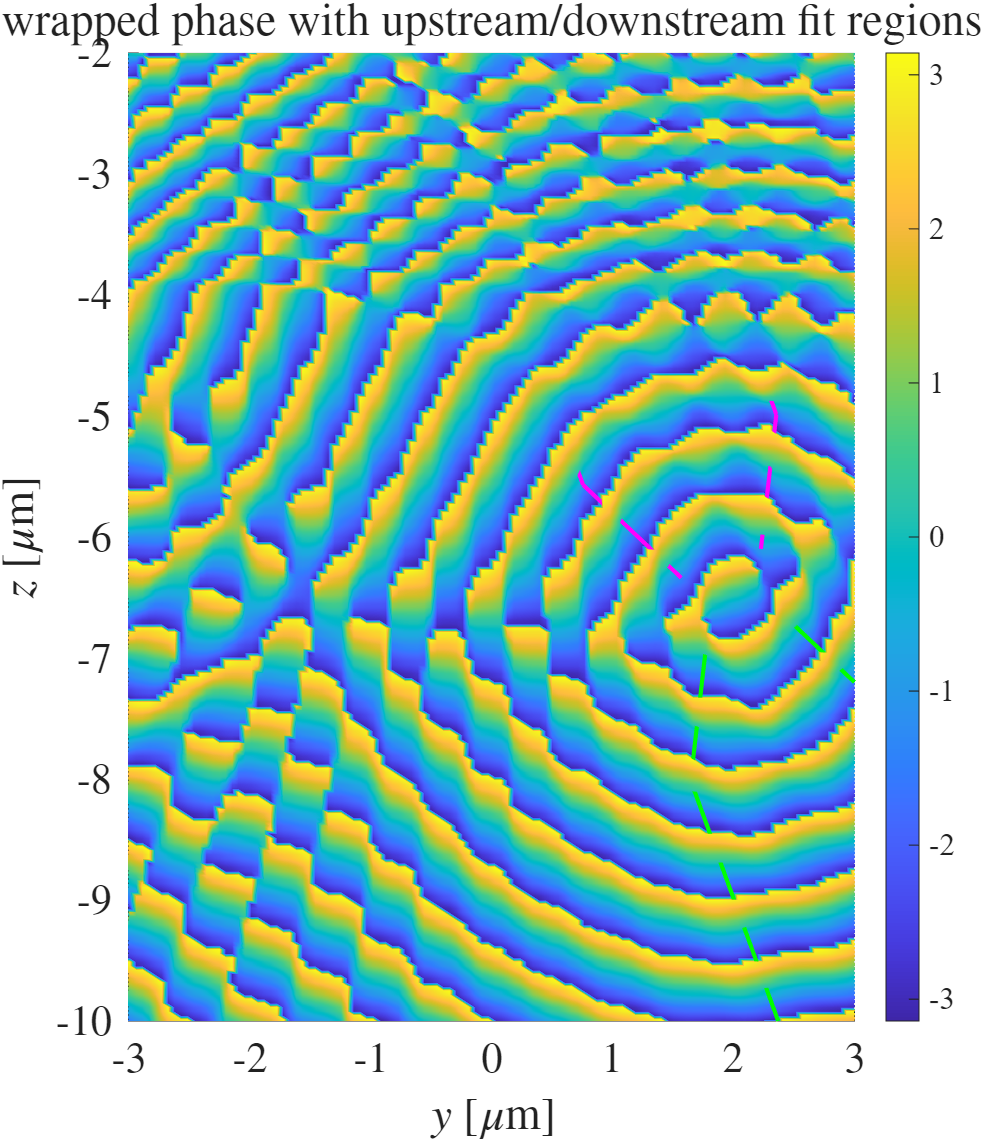}\includegraphics[width=0.5\linewidth]{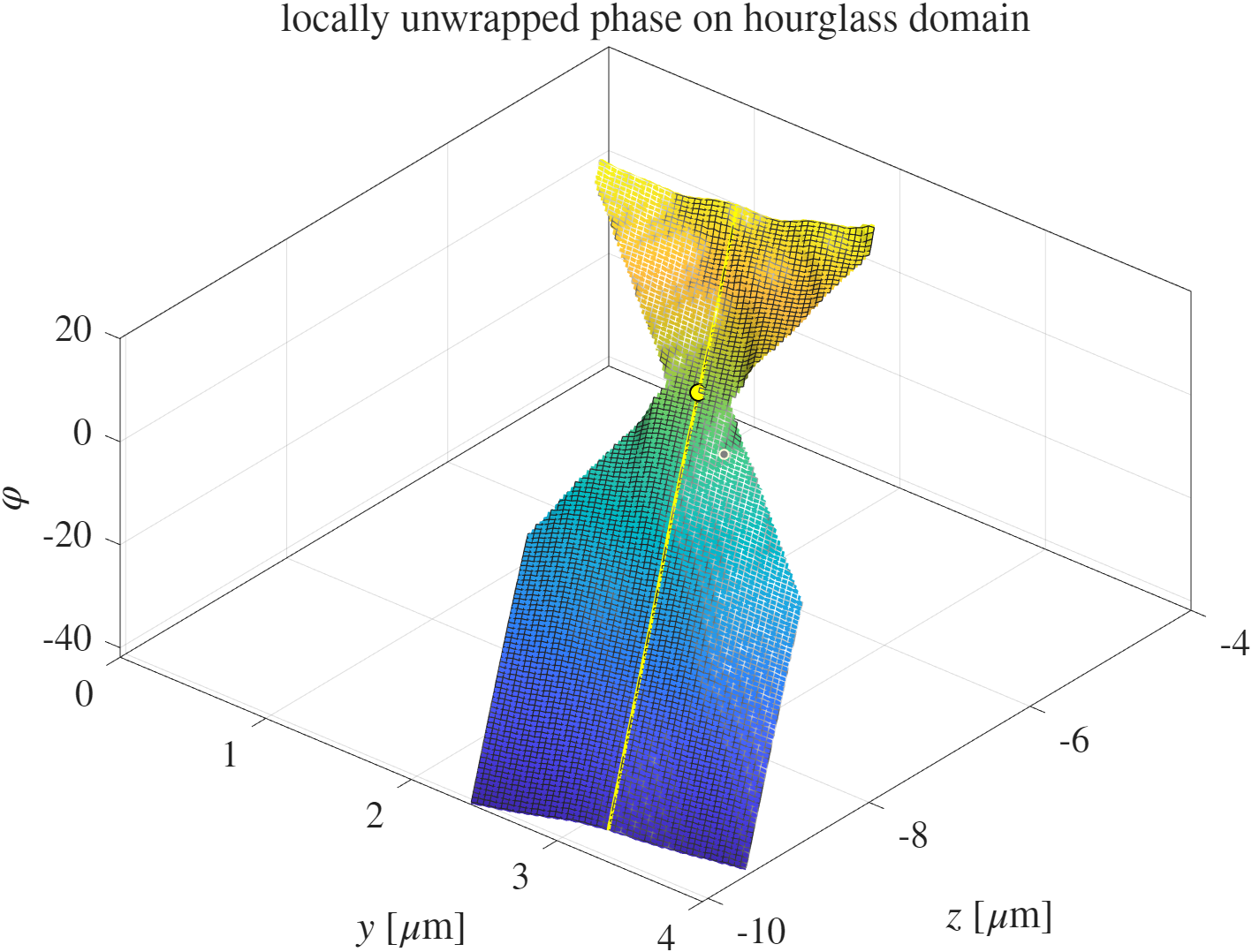}\\\includegraphics[width=0.5\linewidth]{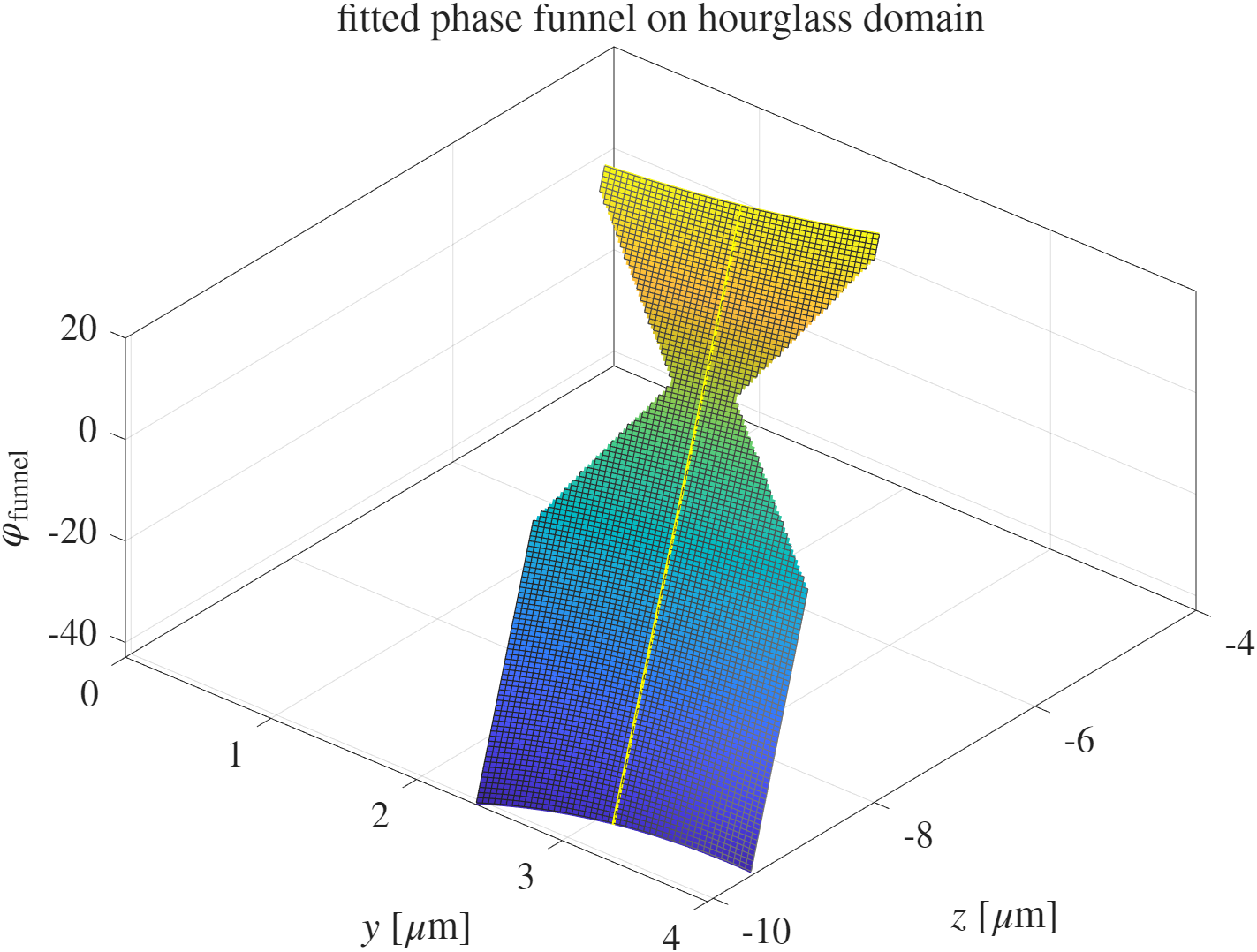}\includegraphics[width=0.35\linewidth]{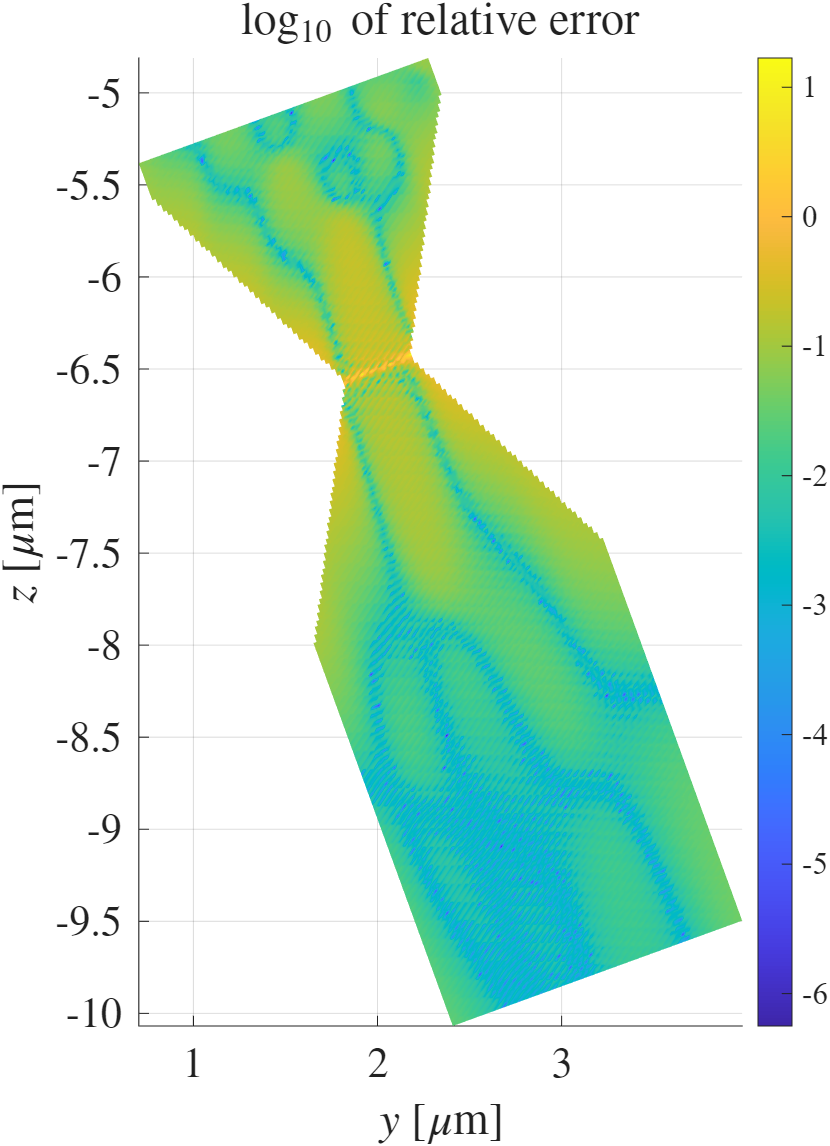}
\caption{Top: Wrapped phase (rad) of PNJ field for the TE$^x$ case, together with an hourglass-shaped domain for phase fitting; unwrapped phase (rad) over the hourglass domain. Bottom: Phase funnel fit and related relative error. Mean relative error over the shown hourglass domain is 4.6\%.}\label{fig:2DrectTE2}
\end{figure}

Consider the time-harmonic scattering of TM$^x$ and TE$^x$, $-z$-directed structured illumination~\cite{2023-phase-only_PNJ,2022-PNJ1,PNJ_CriTob} by a square cross-section dielectric rod, immersed in vacuum, at the free-space operating wavelength $\lambda_0=532$ nm. The scatterer cross-section side length is $8$ $\mu$m, and the refractive index is $n = 1.4607$ (SiO$_2$ at 532 nm). Figures~\ref{fig:2DrectTM} and~\ref{fig:2DrectTE} show example computed PNJ amplitudes $a(y,z)$ and phases $\varphi(y,z)$ for this setup. To diversify our methods, we compute the desired PNJ fields using structured illumination from~\cite{2022-PNJ1} together with the Finite Element package COMSOL Multiphysics~\cite{COMSOL} for the TM$^x$ case and the Python library Ceviche~\cite{CEVICHE} together with time reversal~\cite{PNJ_CriTob} for the TE$^x$ case. For the TM$^x$ case, Figure~\ref{fig:2DrectTM} shows the funnel-like behavior of the unwrapped phase near the PNJ neck, as well as the relative error in approximating the PNJ phase field by the phase funnel expression~\eqref{eqn:phi} in the domain $|y|\le200$ nm, $z\in[-8.5;-4.5]$ $\mu$m. Here, the phase fit parameters are $\varphi_0 = 23.7136$ rad, $k_{\rm eff}  = 10.4777$ $\mu$m$^{-1}$ $\approx$ 0.887 $k_0$, $\Omega = 100$ $\mu$m$^{-3}$, and $z_{\rm PNJ} = -6.5478$ $\mu$m, giving the mean relative error of 1.5\% over the above domain. For the TE$^x$ case, Figure~\ref{fig:2DrectTE} shows the funnel-like behavior of the unwrapped phase near the PNJ neck, and Figure~\ref{fig:2DrectTE2} shows the relative error in approximating the PNJ phase field by the phase funnel expression~\eqref{eqn:ptm} over the shown hourglass domain. The phase fit parameters are $\varphi_0 = 78.2614$ rad, $k_{\rm eff}  = 11.5145$ $\mu$m$^{-1}$ $\approx$ 0.975 $k_0$, and $\Omega = 2.2361$ $\mu$m$^{-3}$, giving the mean relative error of 4.6\% over the shown hourglass domain. For the TE$^x$ fit, we used the fixed known PNJ coordinates $y_{\rm PNJ} = 2$ $\mu$m, $z_{\rm PNJ} = -6.5$ $\mu$m, as well as the fixed known propagation direction $\widehat{\bm d}=(0.3420,-0.9397)$, corresponding to 20$^\circ$ deflection from the $-\widehat{\bm z}$-axis.

Specifically, near the PNJ neck, the phase consistently exhibits a distinct hourglass-like structure, and the mean relative error of the phase funnel model~\eqref{eqn:ptm}, averaged over the two polarizations and PNJ positions, is $3.1\%$. This indicates that the phase funnel model gives an accurate local description near the PNJ neck in these examples.

\begin{figure}[h]
    \centering
    \includegraphics[width=\linewidth]{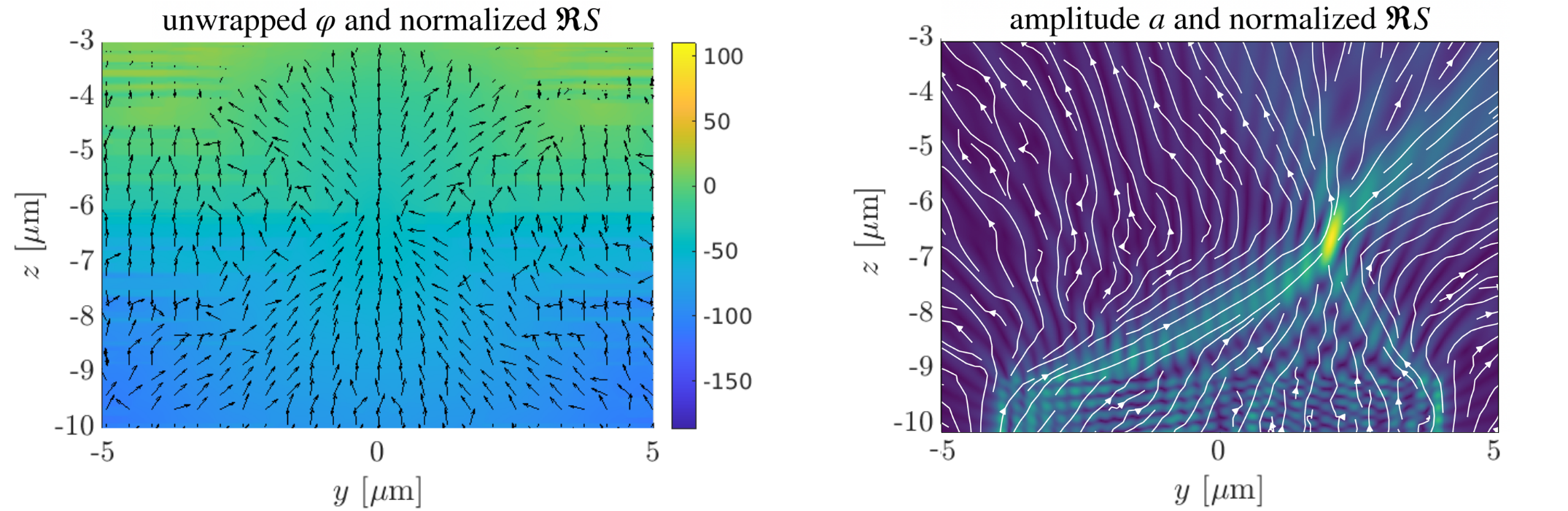}
    \caption{Unwrapped phase and normalized time-average power flow in the PNJ field. Left: 2D TM$^x$ case, right: 2D TE$^x$ case. There is no significant outward transverse ($y$-directed) component of the time-average power flow at the PNJ neck.}
    \label{fig:2D_rect_Re_S}
\end{figure}

Our interpretation of PNJs as effective free-space oscillator structures, which we substantiate in Section~\ref{sec:qqo}, relies furthermore on the facts that 1) PNJ fields exhibit local axial confinement near the waist, and 2) PNJ fields do not exhibit significant local transverse energy leakage away from the jet axis near the jet waist, despite the absence of physical boundaries or material confinement in the surrounding free space. To illustrate the latter point, Figure~\ref{fig:2D_rect_Re_S} shows the PNJ field phase and the normalized real part $\Re\bm S$ of the Poynting vector computed for our 2D numerical examples: we see that there is no significant outward transverse ($y$-directed) component of the time-average power flow at the PNJ neck.

\paragraph{3D case}

\begin{figure}
\centering
\includegraphics[scale=0.85]{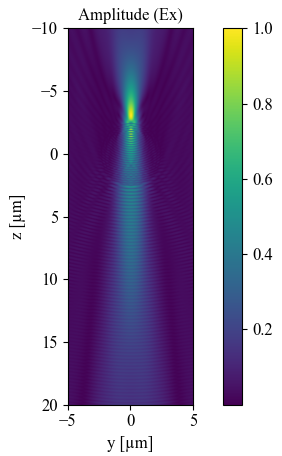}\includegraphics[scale=0.85]{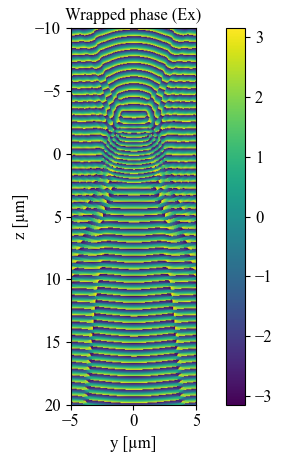}
\caption{A Lorentz-Mie series computation of the amplitude (V/m) and phase (rad) of a TM$^x$-polarized, $-z$-directed 3D PNJ field produced by a spherical micro-element illuminated by a Gaussian beam.}\label{fig:3Dsphe}
\end{figure}

\begin{figure}
\centering
\includegraphics[scale=0.5]{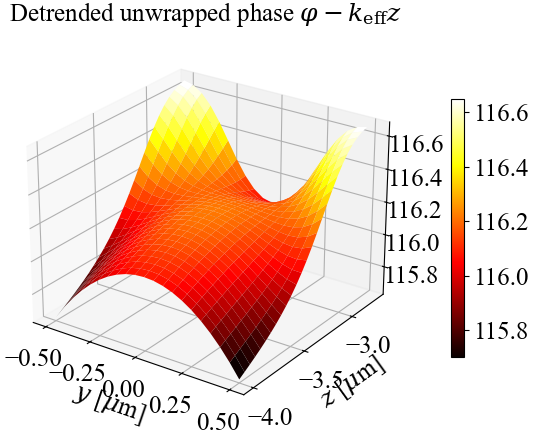}\includegraphics[scale=0.5]{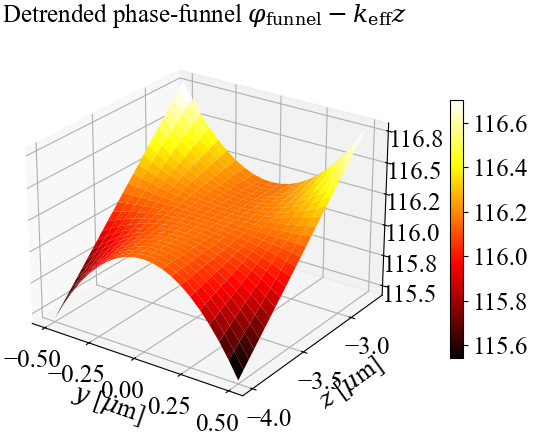}
\includegraphics[scale=0.5]{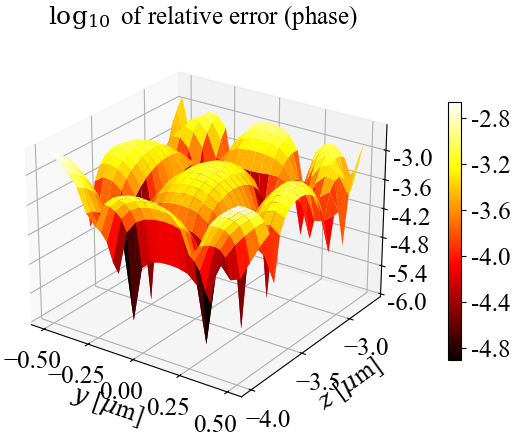}
\caption{3D case: the detrended phase $\varphi(y,z)-k_{\rm eff}z$ (rad), the detrended PNJ phase funnel $\varphi_{\rm funnel}(y,z)-k_{\rm eff}z$ (rad), and the relative error $|\varphi-\varphi_{\rm funnel}|/|\varphi|$. Mean relative error over shown domain is 0.039\%.}\label{fig:3D_phase}
\end{figure}

Consider the time-harmonic scattering of a Gaussian beam, at free-space wavelength $\lambda_0=532$~nm and propagating along the negative $z$-axis, by a dielectric sphere of radius $R=2.5~\mu$m, centered at the origin, with refractive index $n=1.49$, and immersed in vacuum. The Gaussian beam is $\widehat{\bm x}$-polarized, has waist radius $w_0=1.064~\mu$m, coincides with the $z$-axis, and is focused at $z=z_0=2.5~\mu$m. The parameters in this setup follow Huang \emph{et al.}~\cite{Huang}, as well as our paper~\cite{2023-phase-only_PNJ}, where the occurrence of PNJs was documented. Near-field electromagnetic fields were computed using the \texttt{scattnlay} software implementation~\cite{scattnlay}, which realizes generalized Lorenz--Mie theory~\cite{Bobbert1986,ABE} providing exact solutions of Maxwell’s equations for plane wave--illuminated spherical scatterers. To treat arbitrary incidence directions and polarizations, rotational invariance of Maxwell’s equations is exploited: for each incident plane-wave component, the coordinate system is rotated such that the propagation direction aligns with the $z$ axis, the corresponding near field is evaluated, and the resulting electric and magnetic fields are rotated back to the laboratory frame. Gaussian-beam illumination is modeled using an angular-spectrum representation, in which the incident field is expressed as a coherent superposition of plane waves parameterized by polar and azimuthal angles $(\theta,\phi)$ about the $-z$ axis, truncated to a cone $\theta\in[0,\theta_{\max}]$ with $\theta_{\max}=\arcsin(\mathrm{NA})$, where NA$=0.25$. Each component is weighted by a paraxial Gaussian angular spectrum $A(\theta)=\exp\!\left[-(k_0w_0\sin\theta)^2/4\right]$, with an additional phase factor $\exp(jk_0z_0\cos\theta)$ to account for axial displacement of the beam waist reference plane. The angular integrals are approximated numerically using uniform quadrature in $\mu=\cos\theta$ and $\phi$, including the appropriate solid-angle measure. The total field is obtained by coherent summation of all plane-wave contributions. Figure~\ref{fig:3Dsphe} shows the computed total electric field amplitude and phase.
For convenience, and to isolate the transverse phase structure, Figure~\ref{fig:3D_phase} shows the detrended phase $\varphi(y,z)-k_{\rm eff}z$, the detrended PNJ phase funnel $\varphi_{\rm funnel}(y,z)-k_{\rm eff}z$, and the associated relative error $|\varphi-\varphi_{\rm funnel}|/|\varphi|$. The best least-squares fit parameters are computed to be $\varphi_0=116.155$ rad, $k_{\rm eff}=10.9988$ $\mu$m$^{-1}$ $\approx$ 0.931$k_0$, $\Omega = 8.56367$ $\mu$m$^{-3}$, $z_{\rm PNJ}=-3.34$ $\mu$m, and the mean relative error of the fit is 0.039\% within the domain $|y|\le500$ nm, $z\in[-4,-2.7]$ $\mu$m. The very small error reflects the fact that the comparison is restricted to a narrow region near the waist.

\section{PNJs as free-space oscillators}\label{sec:qqo}

We now exploit the observations from Section~\ref{sec:trap} to construct a mesoscopic free-space oscillator PNJ model. Our model holds near the jet axis, for dominant polarization component, and the values of the field confinement parameters depend on the geometry and the refractive index of the PNJ-producing micro-element. However, the structure (hourglass potential leading to harmonic oscillator operator) is robust. The resulting model is local in nature and describes the PNJ as a near-axis, near-waist structure selected by the phase geometry, rather than as a global eigenmode of a bounded cavity. We give numerical examples of our PNJ free-space oscillator modes in Section~\ref{sec:numerical}. In the following we consider time-harmonic electromagnetic fields $(\bm E,\bm H)$ with the time-dependence factor $e^{j\omega t}$ assumed and suppressed.

\paragraph{3D axisymmetric reduction and Laguerre–Gaussian transverse modes.} To justify our approach, which is based in the amplitude-phase decomposition of the Helmholtz equation, assume a dominant linearly ($\widehat{\bm x}$-) polarized electric field component of the photonic nanojet propagating along the negative $z$-axis. Let $\psi = a e^{j\varphi}$ be a complex-valued scalar field satisfying the Helmholtz equation $
(\Delta + k_0^2)\psi = 0$. We introduce an electric Hertz potential vector field $\boldsymbol{\Pi} = \hat{\boldsymbol{x}}\,\psi$, corresponding to a constant polarization direction $\hat{\boldsymbol{x}}$. The associated electromagnetic fields are defined by
\[
\boldsymbol{E} = k_0^2 \boldsymbol{\Pi} + \nabla \operatorname{div}\boldsymbol{\Pi},
\qquad
\boldsymbol{H} = j\omega\varepsilon_0\, \nabla \times \boldsymbol{\Pi}.
\]
The field pair $(\boldsymbol{E},\boldsymbol{H})$ satisfies the source-free time-harmonic Maxwell equations in a homogeneous medium. Indeed,
\[
\nabla \times \boldsymbol{E}
= k_0^2 \nabla \times \boldsymbol{\Pi}
= \frac{\omega^2 \mu_0 \varepsilon_0}{j\omega\varepsilon_0}\,\boldsymbol{H}
= -j\omega\mu_0\,\boldsymbol{H},
\]
and, using the vector identity $\nabla \times \nabla \times = -\Delta + \nabla\operatorname{div}$,
\[
\nabla \times \boldsymbol{H}
= j\omega\varepsilon_0 \nabla \times \nabla \times \boldsymbol{\Pi}
= j\omega\varepsilon_0\big(-\Delta\boldsymbol{\Pi} + \nabla\operatorname{div}\boldsymbol{\Pi}\big)
= j\omega\varepsilon_0\,\boldsymbol{E},
\]
since $\Delta\boldsymbol{\Pi} = \hat{\boldsymbol{x}}\,\Delta\psi = -k_0^2\boldsymbol{\Pi}$. The divergence constraints are automatically satisfied:
\[
\operatorname{div}\boldsymbol{E}
= (\Delta + k_0^2)\partial_x \psi
= \partial_x (\Delta + k_0^2)\psi
= 0,
\qquad
\operatorname{div}\boldsymbol{H}
= j\omega\varepsilon_0\,\operatorname{div}(\nabla\times\boldsymbol{\Pi}) = 0 .
\]
Since $\operatorname{div}\boldsymbol{\Pi} = \partial_x \psi$, the electric field admits the exact representation $\boldsymbol{E}
= k_0^2 \hat{\boldsymbol{x}}\,\psi + \nabla(\partial_x \psi)$. Writing $\psi = a e^{j\varphi}$ and expanding in amplitude--phase variables yields
\[
\boldsymbol{E}
= \psi\Big(
k_0^2 \hat{\boldsymbol{x}}
+ a^{-1}\nabla(\partial_x a)
- (\partial_x\varphi)\,\nabla\varphi
+ j a^{-1}(\partial_x a)\,\nabla\varphi
+ j \nabla(\partial_x\varphi)
+ j a^{-1}(\partial_x\varphi)\,\nabla a
\Big).
\]
In our photonic nanojet setup, the field propagates predominantly along the negative $z$-direction, while remaining strongly polarized along $x$. Accordingly, $
\nabla\varphi \approx -k_0\,\hat{\boldsymbol{z}}$ and $\partial_x \varphi \ll k_0$, with $\partial_x \varphi = 0$ exactly on the jet axis by symmetry. Under this assumption, all terms proportional to $\partial_x\varphi$ are suppressed, and the electric field simplifies to $\boldsymbol{E}
\;\approx\;
\psi\Big(
k_0^2 \hat{\boldsymbol{x}} + a^{-1}\nabla(\partial_x a)
\Big)$. If, in addition, the transverse curvature of the amplitude is weak compared to the wavelength scale, $\|\nabla(\partial_x a)\| \ll k_0^2 a$, or equivalently if the characteristic transverse width $w_x$ satisfies $k_0w_x \gg 1$, then the electric field reduces further to the leading-order form $\boldsymbol{E} \approx k_0^2 \psi\,\hat{\boldsymbol{x}}$. In this regime, the longitudinal field components arise solely from the near-field correction $\nabla(\partial_x\psi)$ and are therefore intrinsically subdominant. We furthermore note that, in the two-dimensional TM$^x$ case, where the electric field vector is out-of-plane and $\partial_xa=0$ as well as $\partial_x\varphi=0$, we have $\bm E=k_0^2\psi\widehat{\bm x}$ exactly. In the two-dimensional TE$^x$ case, we can replace the $\bm E$ field with the $\bm H$ field in the analysis.

The homogeneous Helmholtz equation for $\psi=ae^{j\varphi}$ is equivalent with
\begin{equation}\label{eqn:equiv}
\left\{\begin{array}{rcl}\frac{a}{2}\Delta\varphi+\nabla\varphi\cdot\nabla a&=&0,\\(\Delta+(k_0^2-|\nabla\varphi|^2))a&=&0.
\end{array}\right.
\end{equation}
Inserting~\eqref{eqn:phi} into the second equation in~\eqref{eqn:equiv} gives
\[
\frac{\partial^2a}{\partial r^2}+\frac{1}{r}\frac{\partial a}{\partial r}+\frac{\partial^2a}{\partial z^2}+V(r,z)a=0,
\]
with
\[
V(r,z)=k_0^2-k_{\rm eff}^2-\Omega k_{\rm eff}r^2 + O(r^2(z-z_{\rm PNJ})^2) + O(r^4),\quad\text{near waist and near axis,}
\]
so near the PNJ we may consider the differential equation
\begin{equation}\label{eqn:main_a}
\frac{\partial^2a}{\partial r^2}+\frac{1}{r}\frac{\partial a}{\partial r}+\frac{\partial^2a}{\partial z^2}+(k_0^2-k_{\rm eff}^2)a-\Omega k_{\rm eff}r^2a=0.
\end{equation}
This form expresses the linear variation of transverse phase curvature along the axis, with a vanishing curvature at the waist. Separation of variables $a(r,z)=R(r)Z(z)$ in~\eqref{eqn:main_a} yields the radial form of a 2D quantum harmonic oscillator equation for $R(r)$,
\begin{equation}\label{eqn:R}
R''+\frac{1}{r}R'-\Omega k_{\rm eff}r^2R=(\Lambda-k_0^2+k_{\rm eff}^2)R,
\end{equation}
where $-\Omega k_{\rm eff}r^2$ is the potential term, and $\Lambda-k_0^2+k_{\rm eff}^2$ plays the role of an energy eigenvalue. The solutions of~\eqref{eqn:R} that are bounded at $r=0$ are precisely proportional to the product of a Gaussian and the Kummer $M$-function,
\[
R(r)\propto {\mathrm e}^{-\sqrt{\Omega k_{\rm eff}}r^2/2} M \left(\frac{1}{2}+\frac{\Lambda+k_{\rm eff}^2-k_0^2}{4\sqrt{\Omega k_{\rm eff}}},1,\sqrt{\Omega k_{\rm eff}} r^2\right).
\]
The Kummer $M$-function generally contains a component that grows exponentially with the transverse coordinate. Since we seek a localized near-axis model of the PNJ waist, we exclude this branch and retain only the polynomial branch, obtained when its first parameter is a nonpositive integer, that is, precisely when
\[
\Lambda_n=4\left(n-\frac{1}{2}\right)\sqrt{\Omega k_{\rm eff}}+k_0^2-k_{\rm eff}^2,\quad n\in\{0,-1,-2,\dots\}.
\]
In this case the transverse profile reduces to a Gaussian multiplied by the $|n|$-degree Laguerre polynomial, yielding a confined transverse mode:
\begin{equation}\label{eqn:modes}
R_n(r)\propto e^{-\sqrt{\Omega k_{\rm eff}}r^2/2}L_{|n|}(\sqrt{\Omega k_{\rm eff}}r^2),\quad n\in\{0,-1,-2,\dots\}.
\end{equation}
Importantly, this "quantization" is not imposed by an external boundary condition, but arises as a consequence of the local phase-gradient deficit $|\nabla\varphi|<k_0$, which generates an effective transverse confining potential. The discrete Laguerre–Gaussian modes therefore represent the transversely localized structures selected by the phase geometry of the photonic nanojet in the reduced local model, and not directly imposed by a material structure or a cavity.

\begin{figure}
\centering
\includegraphics[scale=0.6]{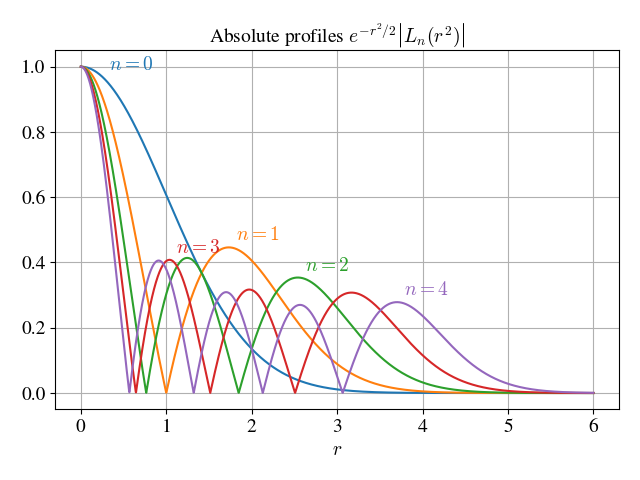}
\caption{Predicted normalized transverse modes of 3D PNJ fields, with the scaling $\Omega k_{\rm eff}=1$.}\label{fig:modes}
\end{figure}

The separation of variables in~\eqref{eqn:main_a} gives $Z''=-\Lambda Z$ in the axial direction. To match the observed local axial confinement of the field near the waist, we require $\Lambda_n>0$, so that the axial factor is oscillatory on a finite waist-centered interval. This yields
\begin{equation}\label{eqn:n}
0\ge n>\frac{1}{2}+\frac{k_{\rm eff}^2-k_0^2}{4\sqrt{\Omega k_{\rm eff}}},
\end{equation}
and allows us to write our general free-space oscillator PNJ model as
\begin{equation}\label{eqn:qqo}
a(r,z)\approx e^{-\sqrt{\Omega k_{\rm eff}}r^2/2}\sum_{n=0}^NL_{n}(\sqrt{\Omega k_{\rm eff}}r^2)\left(A_n\cos\left(\sqrt{\Lambda_{-n}}(z-z_{\rm PNJ})\right)+B_n\sin\left(\sqrt{\Lambda_{-n}}(z-z_{\rm PNJ})\right)\right),
\end{equation}
where
\[
0\le N<\frac{k_0^2-k_{\rm eff}^2}{4\sqrt{\Omega k_{\rm eff}}}-\frac{1}{2}.
\]
In particular, the ground mode ($n=0$) gives the local PNJ profile
\begin{equation}\label{eq:ground3d}
a(r,z)\approx 
\exp\!\left(-\frac{\sqrt{\Omega k_{\rm eff}}}{2}\,r^2\right)
\left(A_0\cos\!\big(\sqrt{\Lambda_0}\,(z-z_{\rm PNJ})\big)+B_0\sin\!\big(\sqrt{\Lambda_0}\,(z-z_{\rm PNJ})\big)\right),
\end{equation}
with $\Lambda_0=k_0^2-k_{\rm eff}^2-2\sqrt{\Omega k_{\rm eff}}$. Condition~\eqref{eqn:n} relates the transverse confinement strength $\Omega$ with the effective axial wavenumber $k_{\rm eff}$ via
\begin{equation}\label{ineq:main}
\Omega<\frac{(k_0^2-k_{\rm eff}^2)^2}{16(N+1/2)^2k_{\rm eff}}\underset{N=0}{\le}\frac{k_{\rm eff}^3}{4}\left(\frac{k_0^2}{k_{\rm eff}^2}-1\right)^2.
\end{equation}
The inequality~\eqref{ineq:main} expresses a trade-off between transverse field confinement (large $\Omega$) and axial radiative power transport (large $k_{\rm eff}$). Specifically, it implies that locally large effective axial wavenumber $k_{\rm eff}\approx k_0$ (or, equivalently, locally small effective axial wavelength $\lambda_{\rm eff}=2\pi/k_{\rm eff}\approx\lambda_0$) cannot support strong transverse field confinement. Thus, near a PNJ waist, we expect $k_{\rm eff}\ll k_0$, that is, $\lambda_{\rm eff}\gg\lambda_0$. A physical interpretation of this trade-off is that the effective axial phase gradient decreases near the PNJ funnel, leading to enhanced local time-average electromagnetic energy density. Furthermore, within the reduced local phase-funnel model, the condition~\eqref{ineq:main} gives lower bounds on FWHM$_a$ and FWHM$_{a^2}$ (the transverse size of PNJ neck) in terms of $\lambda_0$, namely
\begin{equation}\label{ineq:lb1}
{\rm FWHM}_a>\sqrt{\frac{4\ln2}{k_0^2-k_{\rm eff}^2}}=\sqrt{\frac{\ln2}{\pi^2}}\left(\lambda_0^{-2}-\lambda_{\rm eff}^{-2}\right)^{-1/2}\approx0.265\lambda_0(1-\lambda_0^2/\lambda_{\rm eff}^2)^{-1/2}\ge0.265\lambda_0,
\end{equation}
as well as
\begin{equation}\label{ineq:lb2}
{\rm FWHM}_{a^2}>\sqrt{\frac{2\ln2}{k_0^2-k_{\rm eff}^2}}=\sqrt{\frac{\ln2}{2\pi^2}}\left(\lambda_0^{-2}-\lambda_{\rm eff}^{-2}\right)^{-1/2}\approx0.187\lambda_0(1-\lambda_0^2/\lambda_{\rm eff}^2)^{-1/2}\ge0.187\lambda_0,
\end{equation}
where $a({\rm FWHM}_a,0)=0.5a(0,0)$ and $a({\rm FWHM}_{a^2},0)^2=0.5a(0,0)^2$. These bounds are consistent with the observations in the literature that the FWHM decreases with decreasing operating wavelength. Also, they are independent of the position of the PNJ relative to a micro-element, and of the micro-element geometry and material; they rely only on the existence of a PNJ phase funnel as discussed above. The bounds explain the empirical observation that PNJ waists do not collapse arbitrarily below the diffraction scale, even in the near field. Finally, note that each higher-order mode ($n>0$) in~\eqref{eqn:modes} has precisely $|n|$ sidelobes, as illustrated in Figure~\ref{fig:modes}.
This allows more complicated field profiles near the PNJ waist, at the cost of allowing weaker transverse field confinement.

\paragraph{2D Cartesian reduction and Hermite--Gaussian transverse modes.}
In the two-dimensional configuration we work in Cartesian coordinates $(y,z)$, where $-z$ is the axial direction of PNJ propagation and $y$ the transverse coordinate. Inserting the phase funnel ansatz
$
\varphi(y,z)\approx \varphi_0+k_{\rm eff}z+\Omega y^2\,(z-z_{\rm PNJ})/2
$
into the amplitude equation (second line of~\eqref{eqn:equiv}) gives
\begin{equation}\label{eq:ampPDE2d}
\frac{\partial^2a}{\partial y^2}+\frac{\partial^2a}{\partial z^2}+V(y,z)a=0,
\end{equation}
with
\[
V(y,z)=k_0^2-k_{\rm eff}^2-\Omega k_{\rm eff}y^2
+\mathcal{O}\!\left(y^2(z-z_{\rm PNJ})^2\right)+\mathcal{O}(y^4)
\]
near the waist and near the axis, yielding the local free-space oscillator PNJ amplitude model
\begin{equation}\label{eq:main_a_2d}
\frac{\partial^2a}{\partial y^2}+\frac{\partial^2a}{\partial z^2}+(k_0^2-k_{\rm eff}^2)a-\Omega k_{\rm eff}y^2a=0.
\end{equation}
Separating variables $a(y,z)=Y(y)Z(z)$ in~\eqref{eq:main_a_2d} gives
\begin{equation}\label{eq:Y_Z_2d}
Y''-\Omega k_{\rm eff}y^2Y=(\Lambda-k_0^2+k_{\rm eff}^2)\,Y,
\qquad
Z''=-\Lambda Z,
\end{equation}
where $\Lambda$ is the separation constant. The equation for $Y$ in~\eqref{eq:Y_Z_2d} is the one-dimensional quantum harmonic oscillator eigenvalue problem (in Cartesian form). Its solutions can be expressed in terms of parabolic cylinder functions; however, the polynomial branch is obtained precisely when the transverse profile is Hermite--Gaussian, namely
\begin{equation}\label{eq:HG_modes}
Y_n(y)\propto
\exp\!\left(-\frac{\sqrt{\Omega k_{\rm eff}}}{2}\,y^2\right)\,
H_n\!\big((\Omega k_{\rm eff})^{1/4}y\big),
\qquad n=0,1,2,\dots,
\end{equation}
with corresponding discrete separation constants
\begin{equation}\label{eq:Lambda_n_2d}
\Lambda_n
=
k_0^2-k_{\rm eff}^2-(2n+1)\sqrt{\Omega k_{\rm eff}},
\qquad n=0,1,2,\dots.
\end{equation}
As in the cylindrically symmetric case, this ``quantization'' is not imposed by an external boundary condition, but reflects the selection of transversely localized structures compatible with the observed absence of significant outward transverse power flow near the PNJ waist. To enforce the $z$-confinement of the PNJ, we require $\Lambda_n>0$, which leads to
\[
0\le n<\frac{k_0^2-k_{\rm eff}^2}{4\sqrt{\Omega k_{\rm eff}}}-\frac{1}{2},
\]
and allows us to write our general free-space oscillator PNJ model as
\begin{equation}\label{eqn:qqo2d}
a(y,z)\approx e^{-\sqrt{\Omega k_{\rm eff}}y^2/2}\sum_{n=0}^NH_{n}((\Omega k_{\rm eff})^{1/4}y)\left(A_n\cos\left(\sqrt{\Lambda_n}(z-z_{\rm PNJ})\right)+B_n\sin\left(\sqrt{\Lambda_n}(z-z_{\rm PNJ})\right)\right),
\end{equation}
where
\begin{equation}\label{ineq:main2}
0\le N<\frac{k_0^2-k_{\rm eff}^2}{4\sqrt{\Omega k_{\rm eff}}}-\frac{1}{2}.
\end{equation}
In particular, the ground mode ($n=0$) gives the local PNJ profile
\begin{equation}\label{eq:ground2d}
a(y,z)\approx \exp\!\left(-\frac{\sqrt{\Omega k_{\rm eff}}}{2}\,y^2\right)\left(A_0\cos\!\big(\sqrt{\Lambda_0}\,(z-z_{\rm PNJ})\big)+B_0\sin\!\big(\sqrt{\Lambda_0}\,(z-z_{\rm PNJ})\big)\right),
\end{equation}
with $\Lambda_0=k_0^2-k_{\rm eff}^2-\sqrt{\Omega k_{\rm eff}}$.

\begin{figure}
\centering
\includegraphics[scale=0.6]{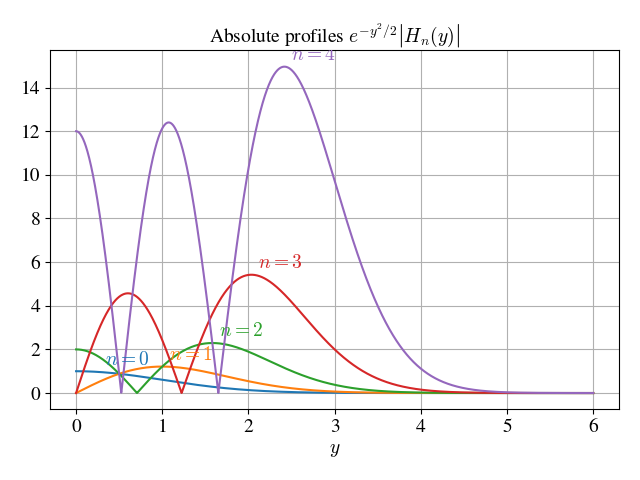}
\caption{Predicted normalized transverse modes of 2D PNJ fields, with the scaling $\Omega k_{\rm eff}=1$.}\label{fig:Hermite}
\end{figure}

Condition~\eqref{ineq:main2} gives precisely the same relation~\eqref{ineq:main} between the transverse field confinement and axial energy transport as in the 3D case; furthermore, the lower bounds on the PNJ FWHM (inequalities~\eqref{ineq:lb1} and~\eqref{ineq:lb2}) from the 3D case analysis apply here as well. Finally, higher-order transverse modes exhibit $n$ sidelobes across $y$, see Figure~\ref{fig:Hermite}, in direct analogy with Hermite--Gaussian beam families, and provide a controlled enlargement of the local model space when the measured PNJ cross section deviates from a pure Gaussian near the waist.

\section{Numerical investigation of the free-space oscillator PNJ model}
\label{sec:numerical}

We now apply the analysis of Section~\ref{sec:qqo} to our numerically computed PNJ fields to verify our free-space oscillator PNJ model and to extract its physical parameters in concrete special cases. The best-fit parameters obtained in this section differ, to varying degrees, from those found in the phase fitting of Section~\ref{sec:trap}. The fitting performed in Section~\ref{sec:trap} is intended to justify the flux-funnel phase model, whereas the fitting carried out here aims to achieve the best possible approximation of the PNJ amplitude profile under the flux-funnel phase ansatz.

\begin{figure}
\centering
\includegraphics[scale=0.3]{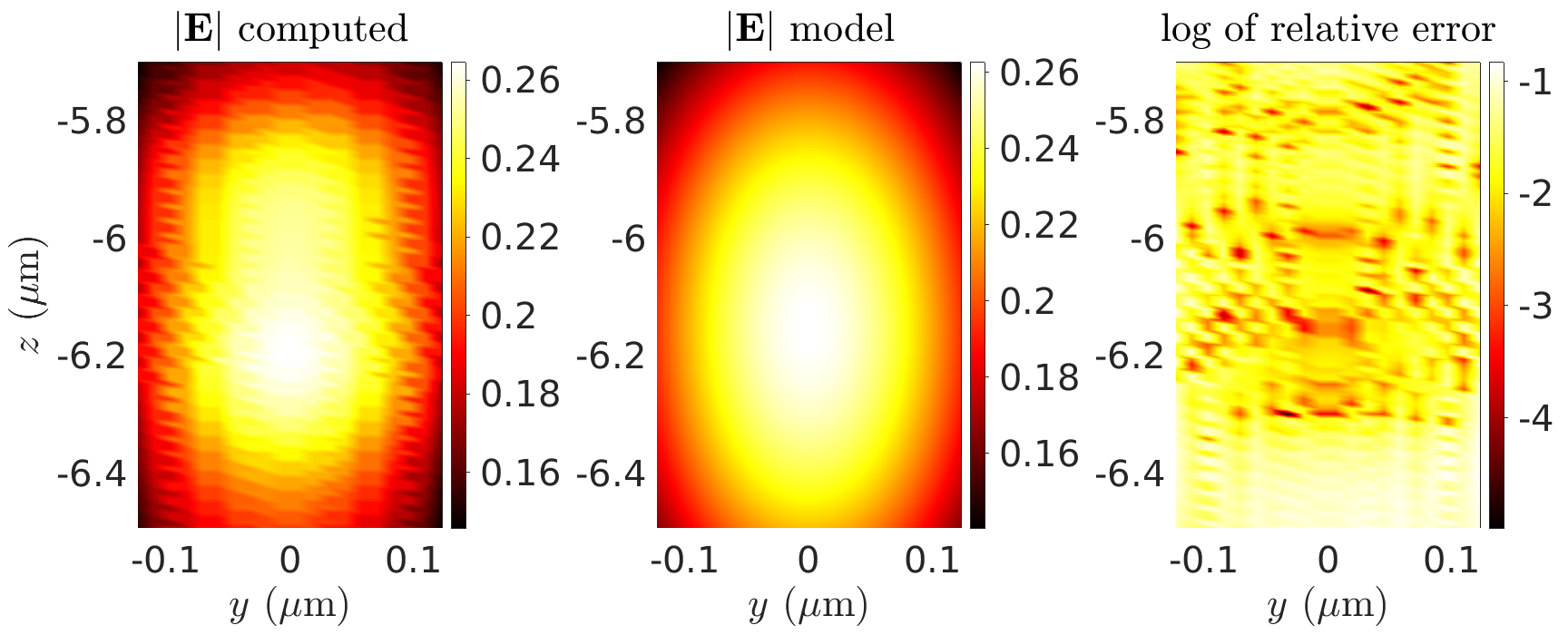}
\caption{Fitting the free-space oscillator PNJ model of Section~\ref{sec:qqo} in the 2D TM$^x$ case with square cross-section micro-element. Mean relative error 3.53\%.}\label{fig:2D_PNJ_1}
\end{figure}

\paragraph{2D TM$^x$ case.}
We model the PNJ field amplitude locally by the ground Hermite-Gaussian mode ($n=0$)~\eqref{eq:ground2d}, where we set $B_0=0$ for simplicity. The COMSOL-computed (and interpolated to a finer grid) amplitude is fitted in MATLAB using a centered and scaled polynomial
\begin{equation}
p(y,z)=\sum_{i+j\le2} p_{ij}
\left(\frac{y}{\sigma_y}\right)^i
\left(\frac{z-\mu_z}{\sigma_z}\right)^j ,
\label{eq:polyfit}
\end{equation}
with scaling parameters
\(\sigma_y = 7.439\times10^{-8}\,\mathrm{m}\),
\(\sigma_z = 2.313\times10^{-7}\,\mathrm{m}\),
and
\(\mu_z = -6.097\times10^{-6}\,\mathrm{m}\).
Table~\ref{tab:1} lists the fitted coefficients \(p_{ij}\) with $95\%$ confidence intervals.
\begin{table}[h]
\centering
\begin{tabular}{lcc}
\toprule
Coefficient & Estimate & 95\% CI \\
\midrule
$p_{00}$ & $0.2616$     & $(0.2610,\; 0.2622)$\\
$p_{10}$ & $0.0003808$  & $( -0.0007756,\; 0.001537 )$ \\
$p_{01}$ & $-0.009893$  & $( -0.01103,\; -0.008755 )$ \\
$p_{20}$ & $-0.02876$   & $( -0.02966,\; -0.02786 )$ \\
$p_{11}$ & $0.001135$   & $( 0.0004064,\; 0.001864 )$ \\
$p_{02}$ & $-0.02012$   & $( -0.02100,\; -0.01924 )$ \\
\bottomrule
\end{tabular}
\caption{Quadratic polynomial fit coefficients with $95\%$ confidence intervals.}
\label{tab:1}
\end{table}
Matching the Taylor expansion of~\eqref{eq:ground2d} up to second order in $y$ and $(z-z_{\rm PNJ})$ with~\eqref{eq:polyfit} gives the closed-form estimates
\[
z_{\rm PNJ}=\mu_z-\frac{p_{01}}{2p_{02}}\,\sigma_z,\qquad
A_0=p_{00}-\frac{p_{01}^2}{4p_{02}},\qquad
\Lambda_0=-\frac{2p_{02}}{A_0}\,\sigma_z^{-2},\qquad
\sqrt{\Omega k_{\rm eff}}=-\frac{2p_{20}}{p_{00}}\,\sigma_y^{-2}.
\]
Numerically, we have $A_0=0.2628$, $z_{\rm PNJ}=-6.1539~\mu\mathrm{m}$, $\Lambda_0=2.862\times10^{12}~\mathrm{m}^{-2}$, and $\sqrt{\Omega k_{\rm eff}}=3.973\times10^{13}~\mathrm{m}^{-2}$. With $\lambda_0=532~\mathrm{nm}$ and $k_0=2\pi/\lambda_0$, the 2D ground-mode relation in~\eqref{eq:ground2d}, $\Lambda_0=k_0^2-k_{\rm eff}^2-\sqrt{\Omega k_{\rm eff}}$, yields $k_{\rm eff}=9.84\times10^{6}~\mathrm{m}^{-1}=9.84~\mu\mathrm{m}^{-1}$ and $\Omega=\frac{(\sqrt{\Omega k_{\rm eff}})^2}{k_{\rm eff}}
=1.60\times10^{20}~\mathrm{m}^{-3}
=160~\mu\mathrm{m}^{-3}$. Figure~\ref{fig:2D_PNJ_1} shows the actual computed PNJ field amplitude, our fitted model field amplitude, and the relative error. Over the shown domain, the mean relative error is 3.53$\%$.
Uncertainties are propagated from the reported coefficient confidence intervals using first-order (Gaussian) error propagation, interpreting the $95\%$ intervals as $\pm 1.96\sigma$ and neglecting coefficient correlations. This gives $A_0=(0.2628\pm0.0003)$, $z_{\rm PNJ} = (-6.154 \pm 0.004)\,\mu\mathrm{m}$, $\Lambda_0 = (2.86 \pm 0.06)\times10^{12}\,\mathrm{m}^{-2}$, $k_{\rm eff} = (9.84 \pm 0.03)\,\mu\mathrm{m}^{-1}$ (for reference, $k_0=11.8105$ $\mu$m$^{-1}$), and $\Omega = (1.60 \pm 0.06)\times10^{20}\,\mathrm{m}^{-3} = (160 \pm 6)\,\mu\mathrm{m}^{-3}$. These values are consistent with the expected local structure of the ground-mode model and with the observed robustness of the PNJ waist in the numerical field data. We note that the effective axial wavenumber is approximately $83\%$ of the free-space value.

\begin{figure}
\centering
\includegraphics[scale=0.55]{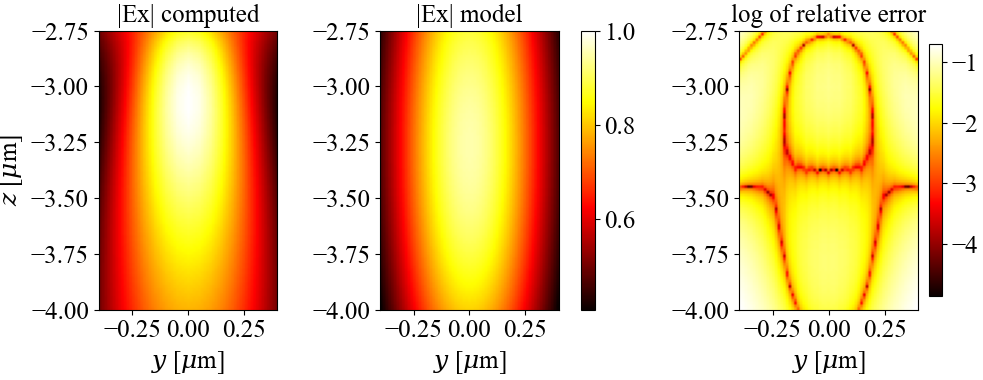}
\caption{Fitting the free-space oscillator PNJ model of Section~\ref{sec:qqo} in the 3D case with spherical micro-element. Mean relative error 3.52\%.}\label{fig:3D_PNJ}
\end{figure}

\paragraph{3D case.} Figure~\ref{fig:3D_PNJ} shows fitting of the ground mode~\eqref{eq:ground3d} against the Lorentz-Mie computed amplitude of the $x$-component of the PNJ electric field. We find the least-squares fit parameter values $A_0=0.896150$, $B_0=0.317225$, $\Omega=6.695850$ $\mu$m$^{-3}$, $k_{\rm eff}=11.033629$ $\mu$m$^{-1}$ $\approx$ $0.934k_0$, $z_{\rm PNJ}=-3.72$ $\mu$m, and $\Lambda_0=0.556267$ $\mu$m$^{-2}$. The mean relative error over the shown domain ($|y|\le400$ nm, $z\in[-4,-2.7]$ $\mu$m) is 3.52\%.

\section{Conclusion}\label{sec:conclusion}

We showed that photonic nanojets exhibit a robust local "PNJ phase funnel" in the waist region, well approximated by \eqref{eqn:ptm}, in which the transverse phase curvature varies linearly with the axial coordinate and vanishes at the waist. Substituting this ansatz into the Helmholtz amplitude equation produces an effective transverse confining potential and a free-space oscillator structure. Restricting attention to bounded, transversely localized profiles in the reduced local model leads to discrete transverse mode families: Laguerre-Gaussian modes in the axisymmetric 3D reduction and Hermite-Gaussian modes in the 2D reduction. The resulting separation constants imply a quantitative trade-off~\eqref{ineq:main} between transverse confinement and axial radiative transport, yielding the lower bounds \eqref{ineq:lb1}-\eqref{ineq:lb2} on PNJ waist widths. Numerical experiments in representative 2D and 3D configurations support the phase funnel model and show that the ground-mode approximation provides a compact local description of the computed PNJ amplitude near the waist. Together, these results support the interpretation of PNJs as mesoscopic transverse modes in free space and provide a parameterized bridge from full-wave fields to a reduced, physically interpretable model.

\section*{Funding}
This work was supported by the Villum Foundation research grant no. 58857.

\end{document}